\documentclass[conference]{IEEEtran}
\IEEEoverridecommandlockouts
\usepackage{cite}
\usepackage{amsmath,amssymb,amsfonts}
\usepackage{algorithmic}
\usepackage{graphicx}
\usepackage{textcomp}
\usepackage{xcolor}
\def\BibTeX{{\rm B\kern-.05em{\sc i\kern-.025em b}\kern-.08em
    T\kern-.1667em\lower.7ex\hbox{E}\kern-.125emX}}
\begin{document}

\title{Performance Analysis of Post-Training Quantization for CNN-based Conjunctival Pallor Anemia Detection\\
\thanks{NSF-EPSCoR Center for the Advancement of Wearable Technologies [NSF Award OIA-1849243]}
}


\author{
Sebastián A. Cruz Romero$^{1}$,
Wilfredo E. Lugo Beauchamp$^{2}$ \\
\textit{Computer Science and Engineering},\\
\textit{University of Puerto Rico at Mayagüez}, \\ Mayagüez, Puerto Rico \\
$^{1}$sebastian.cruz6@upr.edu,
$^{2}$wilfredo.lugo1@upr.edu
}

\maketitle

\begin{abstract}
Anemia is a widespread global health issue, particularly among young children in low-resource settings. Traditional methods for anemia detection often require expensive equipment and expert knowledge, creating barriers to early and accurate diagnosis. To address these challenges, we explore the use of deep learning models for detecting anemia through conjunctival pallor, focusing on the CP-AnemiC dataset, which includes 710 images from children aged 6–59 months. The dataset is annotated with hemoglobin levels, gender, age, and other demographic data, enabling the development of machine learning models for accurate anemia detection. We use the MobileNet architecture as a backbone, known for its efficiency in mobile and embedded vision applications, and fine-tune our model end-to-end using data augmentation techniques and a cross-validation strategy. Our model implementation achieved an accuracy of 0.9313, a precision of 0.9374, and an F1 score of 0.9773 demonstrating strong performance on the dataset. To optimize the model for deployment on edge devices, we performed post-training quantization, evaluating the impact of different bit-widths (FP32, FP16, INT8, and INT4) on model performance. Preliminary results suggest that while FP16 quantization maintains high accuracy (0.9250), precision (0.9370) and F1 score (0.9377), more aggressive quantization (INT8 and INT4) leads to significant performance degradation. Overall, our study supports further exploration of quantization schemes and hardware optimizations to assess trade-offs between model size, inference time, and diagnostic accuracy in mobile healthcare applications.
\end{abstract}


\section{Introduction}
Anemia is a widespread global health concern that mostly affects women and children in low- and middle-income countries (LMICs); symptoms like exhaustion and weakened immune systems may have an impact on a child's development. \cite{Chaparro2019}\cite{who2023anemia} Standard diagnostic methods for anemia involve measuring blood hemoglobin (Hb) levels, which require specialized equipment and personnel—resources often limited in rural and underserved areas. \cite{Chaparro2019}\cite{GarciaCasal2023} According to the World Health Organization, anemia affects over 40\% of children aged 6 to 59 months and 37\% of pregnant women globally, with the highest rates observed in LMICs with limited access to healthcare. \cite{who2023anemia} Given these challenges, there is a growing interest in non-invasive, portable diagnostic tools for early anemia detection to enable preventive interventions. \cite{AnR2021}\cite{GarciaCasal2023}

One non-invasive approach explores computer-aided diagnostic (CAD) systems capable of analyzing physiological indicators such as changes in pigmentation in conjunctiva pallor as a diagnostic indicator for anemia due to its direct relationship with Hb levels and ease of use. \cite{Sheth1997} Among pallor assessment areas (nail beds, palms, tongue), the conjunctiva is considered particularly sensitive for detecting anemia, due to its minimal intervening tissue layers and direct vascular access. \cite{Sheth1997}\cite{AnR2021} However, the feasibility of conjunctival pallor-based diagnostic tools for anemia detection, especially in resource-limited settings, remains a growing area of research interest without clear consensus on optimal implementation approaches. \cite{Merid2023}

With the advancement of artificial intelligence (AI) and deep learning (DL), various approaches to non-invasive anemia diagnosis have been investigated to address limitations associated with clinical methods. \cite{AnR2021} Prior work introduced machine learning (ML) models to identify anemia status or estimate Hb levels from pallor-based features, though constraints such as reliance on proprietary datasets and lack of data diversity persist. \cite{AnR2021}\cite{CPAnemiC2024} The recently developed CP-AnemiC dataset, which focuses on conjunctival pallor for anemia detection in children, addresses some of these limitations by providing a large, publicly available, and balanced dataset that includes diverse samples from ten regions in Ghana. \cite{CPAnemiC2024}

\begin{figure*}
    \centering
    \includegraphics[width=1\textwidth]{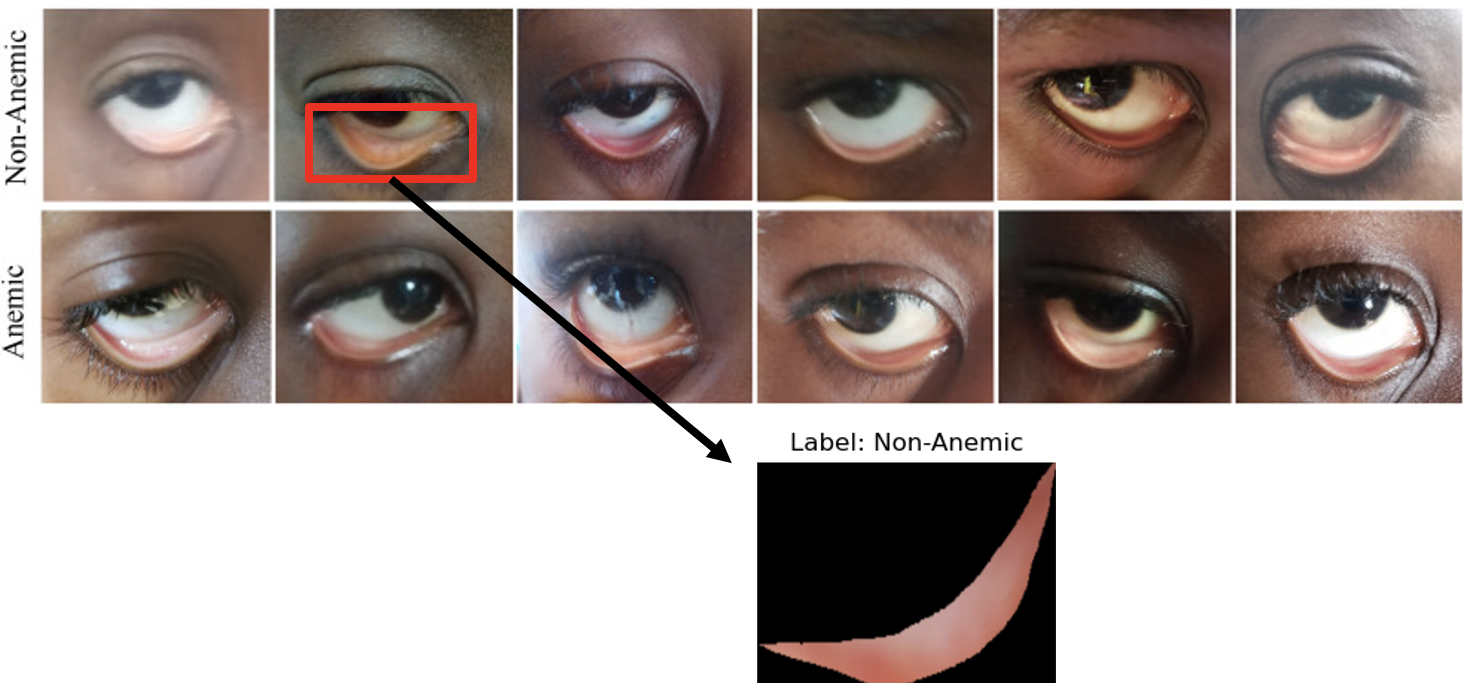}
    \hspace{0.05\textwidth}
    \caption{Sample images of conjunctival pallor with an example region-of-interest from the CP-AnemiC dataset. The first row represents images from the non-anemic patients and the second represents images from the anemic patients.}
    \label{fig:cpanemic-sampleimage}
\end{figure*}

However, challenges persist when introducing models with elevated computational overhead in resource-limited environments. In these scenarios, models designed for high-computation are frequently impracticable, forcing research into lighter, more memory-efficient models that may operate better on edge devices. \cite{Sharma2024} The main disadvantage is that often billions of parameters are used in computation, compared to more traditional algorithms.  \cite{CanzianiPC16} Thus, quantization is introduced as a method to reduce the size of the neural network architecture while maintaining high performance accuracy. Deep learning involves approximating a neural network that uses floating-point numbers with a neural network of reduced bit width representation. Quantization enables reduced memory footprint and computational efficiency by lowering numerical precision, typically from 32-bit floating-point (FP32) to 8-bit integer (INT8). By mapping the original continuous values to a discrete, lower-resolution range, quantization allows substantial gains in storage efficiency and processing speed while operating on reduced precision arithmetic.\cite{Nagel2021}\cite{Gholami2021}.

While this conversion introduces minor approximation errors, several quantization strategies help manage the precision-accuracy trade-off. Post-Training Quantization (PTQ), a widely used method, applies quantization after model training without requiring additional labeled data, making it computationally economical, though at a slight cost to accuracy. In contrast, Quantization-Aware Training (QAT) integrates quantization into the training process, allowing the model to adapt and retain higher accuracy, albeit with increased computational demand during training. \cite{Nagel2021}\cite{Gholami2021} Quantization can follow either uniform or non-uniform schemes; uniform quantization assigns values evenly across intervals, offering simplicity and speed, while non-uniform quantization tailors interval sizes to the data distribution, though at higher computational complexity. PTQ can be applied in static or dynamic forms. Post-training dynamic quantization reduces the bit representation of weights and activations during inference, decreasing computational load and memory usage. This differs from post-training static quantization, which uses a calibration dataset to precompute quantization parameters, including scaling factors for weights and activations, prior to deployment. \cite{Nagel2021}\cite{Gholami2021}. 

However, while certain studies report that quantized models might retain a degree of diagnostic accuracy comparable to full-precision models, more exploration is needed to confirm the consistency of these outcomes in specific diagnostic contexts, such as conjunctival pallor-based anemia detection. In this paper, we introduce a CNN-based classifier with the MobileNet \cite{Howard2017} model for conjunctival pallor-based anemia detection. This project builds on the CP-AnemiC dataset, using it to explore the feasibility of employing PTQ to compare inference performance and execution time at FP32, FP16, INT8, and INT4 bit representations.

\section{Methodology}
\subsection{Dataset}

The CP-AnemiC dataset is a large-scale, publicly available dataset created to address the challenges in anemia detection through conjunctival pallor analysis. It includes 710 conjunctival images from children aged 6–59 months, collected from ten healthcare facilities in Ghana between January and June 2022. Of these, 424 images (60\%) are labeled as anemic and 286 (40\%) as non-anemic, based on the WHO threshold of hemoglobin (Hb) levels below 11 g/dL for anemia diagnosis. This diversity is aimed at enhancing the generalizability of models trained on it. The mean participant age is 31.58 months, with 306 females (43\%) and 404 males (57\%). Each image is annotated with Hb levels, age, gender, collection site, and remarks from laboratory assessments. The dataset also provides demographic analyses of Hb concentration by age and gender, highlighting lower Hb levels in anemic participants.

\begin{table}[h!]
\centering
\caption{A patient-level characteristics summary of the dataset.}
\begin{tabular}{p{3cm}p{1cm}p{1cm}p{1cm}}
\hline
\textbf{Patient Class} & \textbf{Anemic} & \textbf{Non-anemic} & \textbf{Total} \\ \hline
Patients & 424 & 286 & 710 \\
Female & 174 & 132 & 306 \\
Male & 250 & 154 & 404 \\
Age (months) & 31.04 $\pm$ 17.02 & 32.31 $\pm$ 16.46 & 31.58 $\pm$ 16.78 \\ \hline
\multicolumn{4}{c}{\textbf{Anemia Diagnosis for Age 6–59 months}} \\ \hline
Anemia Classification & Anemic & Non-anemic & \\
Hemoglobin Levels & $<$11 g/dL & $\geq$ 11 g/dL & \\ \hline
\end{tabular}
\end{table}

\subsection{Experimental Setting}

\begin{figure*}[ht!]
    \centering
    \includegraphics[width=1\textwidth]{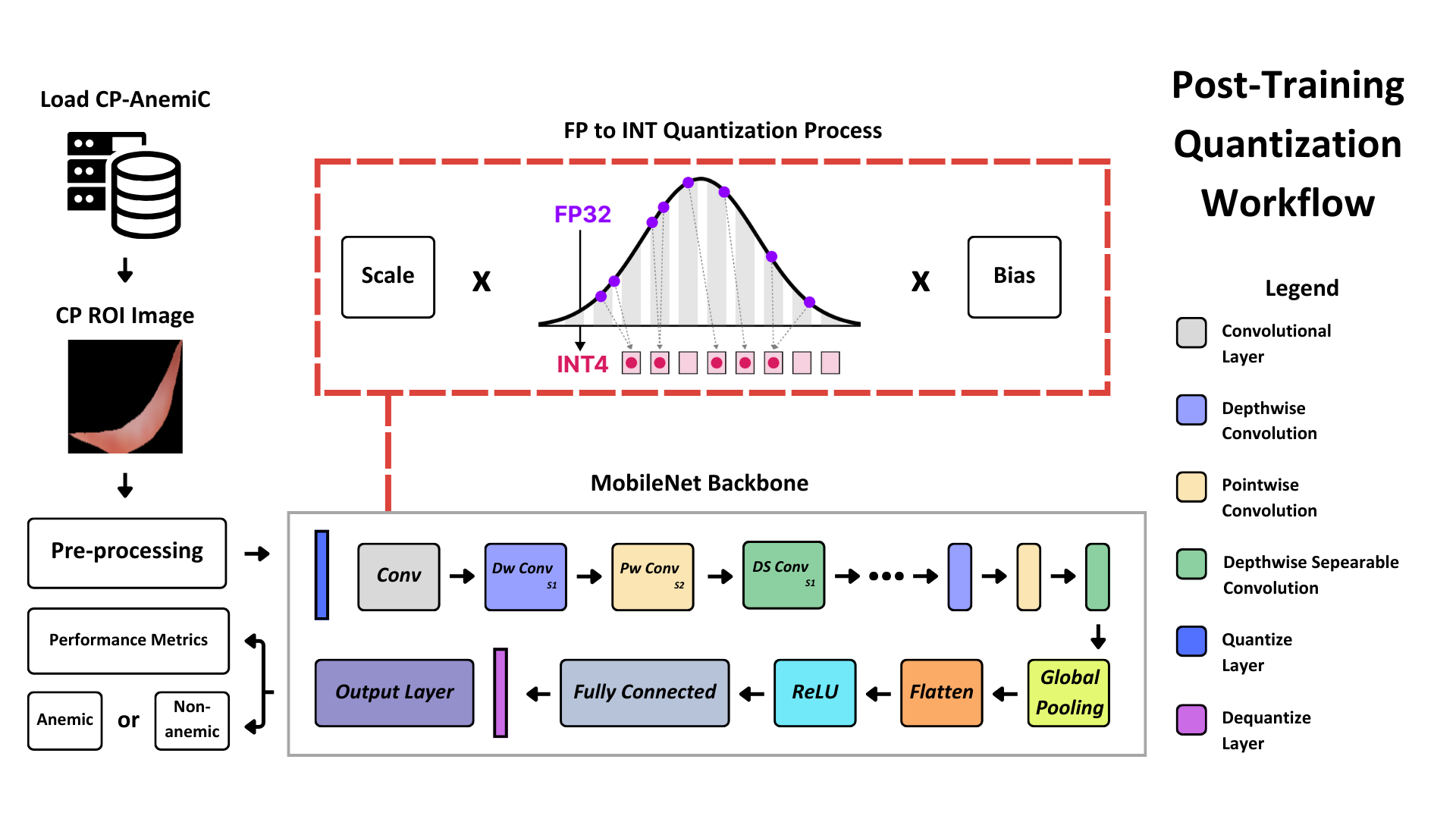}
    \hspace{0.05\textwidth}
    \caption{MobileNet architecture used as a backbone for our anemia detection task. A conjunctival pallor region-of-interest is used as input and the model outputs a numerical representation of anemic and non-anemic classes. The model is quantized for certain layers to the bit-width representation that is selected.}
    \label{fig:mobilenet_framework}
\end{figure*}

We utilize the MobileNet \cite{Howard2017} architecture as the backbone for our anemia classification model, fine-tuning the weights from a pre-trained model on the ImageNet \cite{deng2009imagenet} dataset as illustrated in Figure 2. Our experimental setup replicates the approach described by Appiahene et al. \cite{CPAnemiC2024} for CNN-based anemia detection, including data augmentation techniques such as random horizontal flip, random rotation, random shifts, and random scaling to improve model generalization.

A 5-fold cross-validation strategy was implemented, where four folds were used for training and one for testing. The model weights were randomly initialized at the start of each fold, and training was performed with a batch size of 32 on an NVIDIA GeForce RTX 4090 GPU. We used Binary Cross-Entropy as the loss function and optimized the model with the Adam optimizer, setting the learning rate to \(10^{-4}\). The fully connected layer was modified to output a binary classification target, where 1 represents an anemic class and 0 represents a non-anemic class, using a sigmoid activation function for the output. Training was conducted end-to-end for a maximum of 150 epochs, with early stopping applied if there was no improvement in F1 score for 10 consecutive epochs during the validation step. We compute model evaluation performance metrics as follows:

\begin{align}
    \text{Accuracy} &= \frac{\text{tp} + \text{tn}}{\text{tp} + \text{tn} + \text{fp} + \text{fn}} \\
    \text{Precision} &= \frac{\text{tp}}{\text{tp} + \text{tp}} \\
    \text{Recall} &= \frac{\text{tp}}{\text{tp} + \text{fn}} \\
    \text{F1 Score} &= \frac{2 \times \text{Precision} \times \text{Recall}}{\text{Precision} + \text{Recall}} \\
    \text{AUC} &= \int_{0}^{1} \text{tp rate} \, d(\text{fp rate})
\end{align}

where tp, tn, fp, and fn represent true positives, true negatives, false positives, and false negatives, respectively.

\subsection{Quantization and Inference}

We perform post-training quantization (PTQ) to optimize our fine-tuned MobileNetV2 model by reducing the bit-width of model weights and activations. The model weights are converted into the Open Neural Network Exchange (ONNX) \cite{onnxruntime} format, a universal representation that ensures compatibility across frameworks and optimization tools. This conversion involves exporting the PyTorch model and mapping functional operators to ONNX equivalents while preserving the computational graph. ONNX facilitates backend-specific optimizations by standardizing model representation. In our workflow, NVIDIA's ModelOpt \cite{modelopts} offers quantization configurations with TensorRT \cite{tensorrt} as the backend to dynamically select the optimal precision for operators in each layer based on computational efficiency and accuracy requirements. TensorRT applies these optimizations layer-by-layer, leveraging Tensor Cores for FP16 operations and integer arithmetic units for INT8 and INT4 operations. Dequantization is performed as needed to ensure compatibility between mixed-precision layers. For FP16 conversion, each 32-bit floating-point value $(x)$ is truncated to 16 bits by retaining a reduced number of mantissa bits, represented as:
\[
\text{FP16 Value} = \text{Round}\left(\frac{x}{2^{k}}\right) \cdot 2^{k}
\]
where,
\begin{itemize}
    \item $k$ determines the precision range.
\end{itemize}

For INT8 and INT4 quantization, activations are mapped to integer ranges using scale factors derived from calibration data. We used ModelOpts INT8 default configuration to enable 8-bit precision for weights and activations. Weight quantization is performed per-channel, while activation quantization adopts a per-tensor approach. The MaxCalibrator algorithm determines scale factors by computing the maximum absolute value across tensors, ensuring robust mapping from FP32 to INT8. This configuration prioritizes minimal accuracy degradation while achieving significant computational efficiency.

\[
q_x = \text{Round}\left(\frac{x}{s_x}\right), \quad
s_x = \frac{\max(|x|)}{2^{b-1}-1}
\]

where,
\begin{itemize}
    \item \textit{x} represents the tensor (weights or activations),
    \item \textit{b} is the number of bits (e.g., \textit{b}=8 for INT8 or \textit{b}=4 for INT4).
\end{itemize}

For INT4 quantization, we employed the Advanced Weight Quantization (AWQ) configuration, which utilizes block-wise quantization for weights with block sizes of 128 elements. Activations are excluded from quantization to reduce precision loss. AWQ is used with the "awq\_lite" method, iteratively adjusting scaling parameters ($\alpha$) in small steps to minimize quantization error. This approach achieves extreme compression, targeting environments with strict memory and computational constraints.

\[
\alpha^{(t+1)} = \alpha^{(t)} - \eta \cdot \nabla_{\alpha} \mathcal{L}(w, q_w),
\]

where,
\begin{itemize}
    \item $\eta$ is the step size,
    \item $\mathcal{L}(w, q_w)$ is the loss function measuring quantization error.
\end{itemize}

\section{Results}
\subsection{Anemia Detection}

The model was initialized to train over 150 epochs through 5-fold cross validation due to the limited dataset size. Validation F1 score was monitored to save model weights optimal for running inference at different bit-widths. The highest validation F1 score of 0.9773 was observed at 26 epochs with a validation accuracy of 96.88\% and precision of 97.13\% as observed in Table III. However, to avoid overfitting, early stopping was triggered after 26 epochs indicating model convergence as further F1 score improvements were minimal.

\begin{table}[h!]
\centering
\caption{Anemia Classification Training Performance}
\begin{tabular}{p{.25cm}p{1cm}p{1cm}p{1cm}p{1cm}p{1cm}p{1cm}}
\hline
Fold & \textbf{Loss} & \textbf{Accuracy} & \textbf{Precision} & \textbf{Recall} & \textbf{F1 Score} & \textbf{AUC Score} \\
\hline
74  & 0.2269 & 0.9229 & 0.9252 & 0.9531 & 0.9383 & 0.9667 \\
98  & 0.2157 & 0.9208 & 0.9305 & 0.9402 & 0.9331 & 0.9680 \\
100 & 0.2448 & 0.9160 & 0.9370 & 0.9267 & 0.9257 & 0.9636 \\
99  & 0.2314 & 0.9122 & 0.8869 & 0.9372 & 0.9090 & 0.9759 \\
80  & 0.2090 & 0.9104 & 0.9170 & 0.9293 & 0.9219 & 0.9705 \\
103 & 0.2023 & 0.9092 & 0.9312 & 0.9229 & 0.9249 & 0.9755 \\
86  & 0.2394 & 0.9090 & 0.9160 & 0.9271 & 0.9208 & 0.9710 \\
101 & 0.2230 & 0.9076 & 0.9598 & 0.8934 & 0.9225 & 0.9767 \\
71  & 0.2663 & 0.9021 & 0.9071 & 0.9307 & 0.9176 & 0.9519 \\
78  & 0.2495 & 0.9000 & 0.9244 & 0.9155 & 0.9166 & 0.9573 \\
91  & 0.2521 & 0.8988 & 0.9085 & 0.9252 & 0.9132 & 0.9614 \\
79  & 0.2356 & 0.8979 & 0.9046 & 0.9286 & 0.9127 & 0.9633 \\
65  & 0.2660 & 0.8972 & 0.9329 & 0.8979 & 0.9115 & 0.9591 \\
76  & 0.2808 & 0.8958 & 0.9107 & 0.9162 & 0.9072 & 0.9464 \\
85  & 0.2973 & 0.8951 & 0.9047 & 0.9191 & 0.9061 & 0.9527 \\
92  & 0.2322 & 0.8951 & 0.9057 & 0.9213 & 0.9097 & 0.9677 \\
73  & 0.2531 & 0.8938 & 0.9204 & 0.9063 & 0.9104 & 0.9571 \\
75  & 0.2741 & 0.8935 & 0.9265 & 0.9098 & 0.9147 & 0.9585 \\
84  & 0.2487 & 0.8931 & 0.8993 & 0.9221 & 0.9071 & 0.9624 \\
96  & 0.2725 & 0.8924 & 0.9124 & 0.9144 & 0.9064 & 0.9588 \\
93  & 0.2823 & 0.8903 & 0.9084 & 0.9045 & 0.9053 & 0.9653 \\
89  & 0.3117 & 0.8882 & 0.9086 & 0.9010 & 0.9021 & 0.9559 \\
88  & 0.2817 & 0.8854 & 0.8947 & 0.9079 & 0.8992 & 0.9504 \\
87  & 0.2823 & 0.8851 & 0.8797 & 0.9345 & 0.9001 & 0.9526 \\
82  & 0.2719 & 0.8813 & 0.9103 & 0.8968 & 0.9001 & 0.9502 \\
\hline
\end{tabular}
\end{table}

\begin{table}[h!]
\centering
\caption{Anemia Classification Validation Performance}
\begin{tabular}{p{.25cm}p{1cm}p{1cm}p{1cm}p{1cm}p{1cm}p{1cm}}
\hline
Fold & \textbf{Loss} & \textbf{Accuracy} & \textbf{Precision} & \textbf{Recall} & \textbf{F1 Score} & \textbf{AUC Score} \\
\hline
12  & 0.0857 & 0.9688 & 0.9713 & 0.9722 & 0.9705 & 0.9978 \\
13  & 0.1033 & 0.9688 & 0.9773 & 0.9773 & 0.9773 & 0.9923 \\
11  & 0.1225 & 0.9531 & 0.9565 & 0.9659 & 0.9602 & 0.9920 \\
10  & 0.1273 & 0.9453 & 0.9268 & 0.9747 & 0.9481 & 0.9910 \\
9   & 0.1846 & 0.9315 & 0.8970 & 1.0000 & 0.9453 & 1.0000 \\
8   & 0.3033 & 0.8768 & 0.9147 & 0.8728 & 0.8915 & 0.9497 \\
7   & 0.3598 & 0.8162 & 0.8481 & 0.8302 & 0.8382 & 0.9189 \\
6   & 0.3852 & 0.7858 & 0.7981 & 0.8589 & 0.8239 & 0.9160 \\
5   & 0.4081 & 0.7822 & 0.7570 & 0.9025 & 0.8181 & 0.8923 \\
3   & 0.5210 & 0.7813 & 0.7964 & 0.8160 & 0.7993 & 0.8048 \\
4   & 0.4401 & 0.7537 & 0.7365 & 0.8927 & 0.8017 & 0.8865 \\
1   & 0.5704 & 0.6696 & 0.6731 & 0.8441 & 0.7435 & 0.7548 \\
2   & 0.6277 & 0.6677 & 0.6700 & 0.8615 & 0.7458 & 0.6965 \\
0   & 0.6060 & 0.6530 & 0.6696 & 0.8242 & 0.7292 & 0.6981 \\
\hline
\end{tabular}
\end{table}

\begin{figure*}[ht!]
    \centering
    \includegraphics[width=1\textwidth]{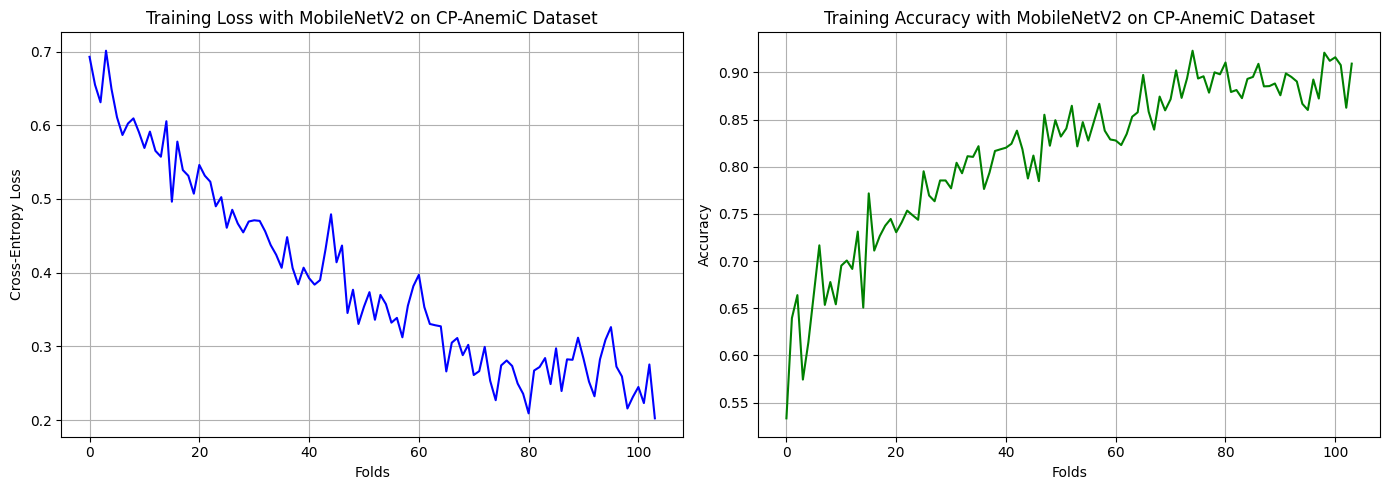}
    \hspace{0.05\textwidth}
    \includegraphics[width=1\textwidth]{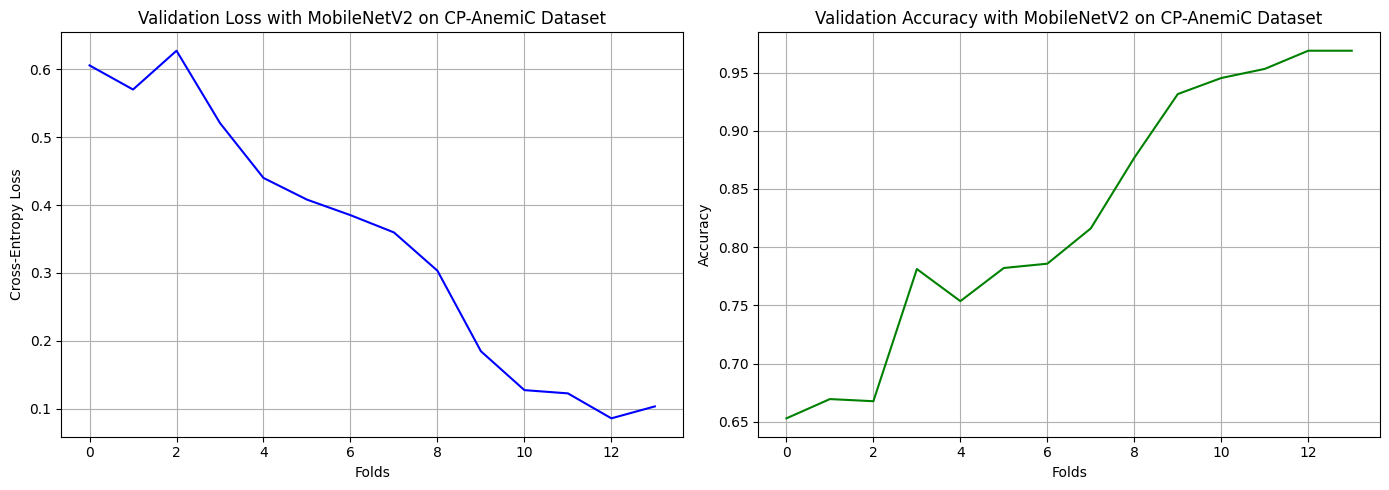}
    \caption{Training and Validation Performance of our fine-tuned MobileNet for Anemia Detection after 26 epochs.}
    \label{fig:model_performance}
\end{figure*}

\subsection{Performance Comparison at Different Quantization Levels}

FP32 achieves the best performance, with a loss of 0.2141, accuracy of 93.13\%, precision of 93.74\%, recall of 95.00\%, F1 score of 0.9428, and AUC score of 0.9657 as shown in Table VI. FP16 follows closely, showing only a slight drop in performance with a loss of 0.2149, accuracy of 92.50\%, and similar precision, recall, and F1 scores, maintaining a strong AUC of 0.9654. In contrast, INT8 quantization leads to a substantial decrease in accuracy (71.25\%) and AUC (90.05\%), with a much higher loss of 0.7441, indicating that the model's performance suffers despite lower computational requirements. The INT4 quantization results in a drastic performance drop, with a loss of 2.3136, accuracy of just 43.13\%, and poor precision (20.00\%) and recall (1.00\%).

\begin{table}[h!]
\centering
\caption{Performance Comparison at Different Quantization Levels}
\begin{tabular}{p{.5cm}p{1cm}p{1cm}p{1cm}p{1cm}p{1cm}p{1cm}}
\hline
\textbf{Bit-width} & \textbf{Loss} & \textbf{Accuracy} & \textbf{Precision} & \textbf{Recall} & \textbf{F1 Score} & \textbf{AUC Score} \\
\hline
FP32 & 0.2141 & 0.9313 & 0.9374 & 0.9500 & 0.9428 & 0.9657 \\
FP16 & 0.2149 & 0.9250 & 0.9370 & 0.9400 & 0.9377 & 0.9654 \\
INT8 & 0.7441 & 0.7125 & 0.7697 & 0.7519 & 0.7607 & 0.9005 \\
INT4 & 2.3136 & 0.4313 & 0.2000 & 0.0100 & 0.0196 & 0.6387 \\
\hline
\end{tabular}
\end{table}

\subsection{Quantized Layers for Integer Arithmetic}

Table IV demonstrate trade-offs in memory and execution time across quantization levels. FP16 achieves significant memory reduction (4.61 MB) and the fastest execution time (37.4 ms). INT8, despite its reduced precision, shows increased model size (9.24 MB) and latency (91.9 ms), likely due to overhead from quantization parameters layered a top of existing architecture. INT4, while achieving extreme compression, results in the largest model size (17.75 MB) and comparable latency to FP32 (49.5 ms), indicating inefficiencies in handling ultra-low precision. Table  V summarizes the key layers affected by INT8 and INT4 quantization, highlighting their weight quantization properties.
\subsection{Quantized Layers for Integer Arithmetic}

Table \ref{tab:mem_con} demonstrate trade-offs in memory and execution time across quantization levels. FP16 achieves significant memory reduction (4.61 MB) and the fastest execution time (37.4 ms). INT8, despite its reduced precision, shows increased model size (9.24 MB) and latency (91.9 ms), likely due to overhead from quantization parameters layered a top of existing architecture. INT4, while achieving extreme compression, results in the largest model size (17.75 MB) and comparable latency to FP32 (49.5 ms), indicating inefficiencies in handling ultra-low precision. Table  V summarizes the key layers affected by INT8 and INT4 quantization, highlighting their weight quantization properties.

\begin{table}[h!]
\centering
\caption{Quantization Summary for Key Layers}
\begin{tabular}{p{2cm}p{.5cm}p{1.5cm}p{2cm}}
\hline
\textbf{Layer Name} & \textbf{Bit-width} & \textbf{Quantization Method} & \textbf{amax Range} \\
\hline
features.0.0 & INT8 & Per-axis & [0.0039, 1.4840] \\
features.1.conv.0.0 & INT8 & Per-axis & [0.0036, 2.6928] \\
features.10.conv.1.0 & INT8 & Per-axis & [0.0103, 0.6126] \\
features.0.0 & INT4 & Block-wise & [0.0005, 1.4840] \\
features.1.conv.0.0 & INT4 & Block-wise & [0.0014, 2.6928] \\
features.10.conv.1.0 & INT4 & Block-wise & [0.0035, 0.5219] \\
\label{tab:layers}

\end{tabular}
\end{table}

\begin{table}[h!]
\centering
\caption{Memory Consumption and Execution Time Across Different Quantization Levels}
\begin{tabular}{p{.5cm}p{2cm}p{2.5cm}}
\hline
\textbf{Bit-width} & \textbf{Model Size} & \textbf{Execution Time} \\
\hline
FP32 & 9.13 MB & 48.6 ms ± 235 $\mu$s \\
FP16 & 4.61 MB & 37.4 ms ± 1.01 ms \\
INT8 & 9.24 MB & 91.9 ms ± 1.92 ms \\
INT4 & 17.75 MB & 49.5 ms ± 1.34 ms \\
\hline
\label{tab:mem_con}
\end{tabular}
\end{table}

\section{Conclusion and Future Work}

This study illustrates the potential of lightweight architectures like MobileNet for anemia detection through conjunctival pallor analysis. Our model achieved state-of-the-art performance on the CP-AnemiC dataset, achieving an F1 score of 0.9428 and an accuracy of 93.13\% during inference without quantization. Post-training quantization further optimized the model for edge device deployment across various quantization levels revealing a nuanced trade-off between computational efficiency and predictive accuracy. FP16 quantization maintained strong performance, with minimal reductions in accuracy (92.50\%) and F1 score (0.9377). Performance degradation was evident with more aggressive quantization techniques. Preliminary results of model performance across different quantization bit-widths show that aggressive quantization can severely degrade the model's predictive capabilities.

Future work includes systematically achieving full integer arithmetic within layer-by-layer computation. Furthermore, the impact of these optimizations on inference latency will be studied, particularly on edge devices such as the NVIDIA Jetson Xavier NX and TX2 NX due to their small form factor and low-power consumption. Using TensorRT for operator conversion backend, we aim to exploit its ability to select optimal precision per layer, leveraging mixed-precision arithmetic to enhance execution speed while maintaining diagnostic integrity.

\section*{Acknowledgment}

This work is supported by the Center for Research \& Development at the University of Puerto Rico at Mayagüez and the NSF-EPSCoR Center for the Advancement of Wearable Technologies, NSF grant OIA-1849243.

\fontsize{9}{9}\selectfont

\end{document}